\documentclass[aps,prl,twocolumn,floats,amsmath,preprintnumbers,nofootinbib,superscriptaddress,amssymb,10pt]{revtex4-2}
\usepackage{graphicx}
\usepackage{dcolumn}
\usepackage{bm,graphicx,hyperref,subfigure}
\usepackage{float,amsfonts,graphicx}
\usepackage{color}
\usepackage{tikz}
\usepackage{adjustbox}
\usetikzlibrary{decorations.pathreplacing}
\usetikzlibrary{arrows.meta}
\usepackage{orcidlink}
\usepackage{ulem}

\usepackage[utf8]{inputenc}
\usepackage{comment}

\usepackage{placeins}
\usepackage{xcolor}

\usepackage{cleveref} 
\crefname{figure}{Figure}{Figures}
\crefname{table}{Table}{Tables}
\crefname{section}{section}{sections} 
\crefname{appendix}{appendix}{appendixes}
\crefname{equation}{Eq.}{Eqs.}
\Crefname{figure}{Figure}{Figures}
\Crefname{table}{Table}{Tables}
\Crefname{section}{Section}{Sections} 
\Crefname{appendix}{Appendix}{Appendixes}
\Crefname{equation}{Eq.}{Eqs.}
\AddToHook{cmd/appendix/before}{\crefalias{section}{appendix}}

\relax

\newcommand{\GF}{G_{\rm{F}}}
\newcommand{\Sew}{S_{\rm{EW}}}
\newcommand{\eq}[1]{\begin{align}\label{#1}}
\newcommand{\en}{\end{align}}
\newcommand{\eqar}[1]{\begin{align}\label{#1}}
\newcommand{\enar}{\end{align}}

\newcommand{\ket}[1]{\ensuremath{| {#1} \rangle }}
\newcommand{\bra}[1]{\ensuremath{\langle {#1} |}}

\newcommand{\remove}[1]{ }

\renewcommand{\vec}[1]{\bm{#1}}

\begin{document}

\setlength\abovedisplayskip{10pt}
\setlength\belowdisplayskip{10pt}

\setlength{\parskip}{14pt}
\setlength{\parindent}{0pt}

\preprint{HIP-2025-10/TH}

\title{Inclusive semileptonic decays of the \texorpdfstring{$D_s$}{Ds} meson:\\ Lattice QCD confronts experiments}

\author{Alessandro~\surname{De~Santis}~\orcidlink{0000-0002-2674-4222}}
\affiliation{Helmholtz-Institut Mainz, Johannes Gutenberg-Universit{\"a}t Mainz, 55099 Mainz, Germany}
\affiliation{GSI Helmholtz Centre for Heavy Ion Research, 64291 Darmstadt, Germany}

\author{Antonio~\surname{Evangelista}~\orcidlink{0000-0002-3320-3176}}
\affiliation{Dipartimento di Fisica \& INFN, Universit\`a di Roma ``Tor Vergata'', Via della Ricerca Scientifica 1, I-00133 Rome, Italy}

\author{Roberto~\surname{Frezzotti}~\orcidlink{0000-0001-5746-0065}}
\affiliation{Dipartimento di Fisica \& INFN, Universit\`a di Roma ``Tor Vergata'', Via della Ricerca Scientifica 1, I-00133 Rome, Italy}

\author{Giuseppe~\surname{Gagliardi}~\orcidlink{0000-0002-4572-864X}}
\affiliation{Dipartimento di Matematica e Fisica, Universit\`a Roma Tre, Via della Vasca Navale 84, I-00146 Rome, Italy}
\affiliation{INFN, Sezione di Roma Tre, Via della Vasca Navale 84, I-00146 Rome, Italy}

\author{Paolo~\surname{Gambino}~\orcidlink{0000-0002-7433-4914}}
\affiliation{Dipartimento di Fisica, Universit\`a di Torino \& INFN, Sezione di Torino, Via Pietro Giuria 1, I-10125 Turin, Italy}

\author{Marco~\surname{Garofalo}~\orcidlink{0000-0002-4508-6421}}
\affiliation{HISKP (Theory) \& Bethe Centre for Theoretical Physics, Rheinische Friedrich-Wilhelms-Universit\"at Bonn, Nussallee 14-16, D-53115 Bonn, Germany}

\author{Christiane~Franziska~\surname{Gro\texorpdfstring{\ss}}~\orcidlink{0009-0009-5876-1455}}
\affiliation{HISKP (Theory) \& Bethe Centre for Theoretical Physics, Rheinische Friedrich-Wilhelms-Universit\"at Bonn, Nussallee 14-16, D-53115 Bonn, Germany}

\author{Bartosz~\surname{Kostrzewa}~\orcidlink{0000-0003-4434-6022}}
\affiliation{HISKP (Theory) \& Bethe Centre for Theoretical Physics, Rheinische Friedrich-Wilhelms-Universit\"at Bonn, Nussallee 14-16, D-53115 Bonn, Germany}

\author{Vittorio~\surname{Lubicz}~\orcidlink{0000-0002-4565-9680}}
\affiliation{Dipartimento di Matematica e Fisica, Universit\`a Roma Tre, Via della Vasca Navale 84, I-00146 Rome, Italy}
\affiliation{INFN, Sezione di Roma Tre, Via della Vasca Navale 84, I-00146 Rome, Italy}

\author{Francesca~\surname{Margari}~\orcidlink{0000-0003-2155-7679}}
\affiliation{Dipartimento di Fisica \& INFN, Universit\`a di Roma ``Tor Vergata'', Via della Ricerca Scientifica 1, I-00133 Rome, Italy}

\author{Marco~\surname{Panero}~\orcidlink{0000-0001-9477-3749}}
\affiliation{Dipartimento di Fisica, Universit\`a di Torino \& INFN, Sezione di Torino, Via Pietro Giuria 1, I-10125 Turin, Italy}
\affiliation{Department of Physics \& Helsinki Institute of Physics, University of Helsinki, PL 64, FIN-00014 Helsinki, Finland}

\author{Francesco~\surname{Sanfilippo}~\orcidlink{0000-0002-1333-745X}}
\affiliation{INFN, Sezione di Roma Tre, Via della Vasca Navale 84, I-00146 Rome, Italy}

\author{Silvano~\surname{Simula}~\orcidlink{0000-0002-5533-6746}}
\affiliation{INFN, Sezione di Roma Tre, Via della Vasca Navale 84, I-00146 Rome, Italy}

\author{Antonio~\surname{Smecca}~\orcidlink{0000-0002-8887-5826}}
\affiliation{Department of Physics, Faculty of Science and Engineering, Swansea University (Singleton Park Campus), Singleton Park, SA2 8PP Swansea, Wales, United Kingdom}

\author{Nazario~\surname{Tantalo}~\orcidlink{0000-0001-5571-7971}}
\affiliation{Dipartimento di Fisica \& INFN, Universit\`a di Roma ``Tor Vergata'', Via della Ricerca Scientifica 1, I-00133 Rome, Italy}

\author{Carsten~\surname{Urbach}~\orcidlink{0000-0003-1412-7582}}
\affiliation{HISKP (Theory) \& Bethe Centre for Theoretical Physics, Rheinische Friedrich-Wilhelms-Universit\"at Bonn, Nussallee 14-16, D-53115 Bonn, Germany}

\begin{abstract}
\vspace{0.05cm}
\centerline{\includegraphics[height=5.3cm]{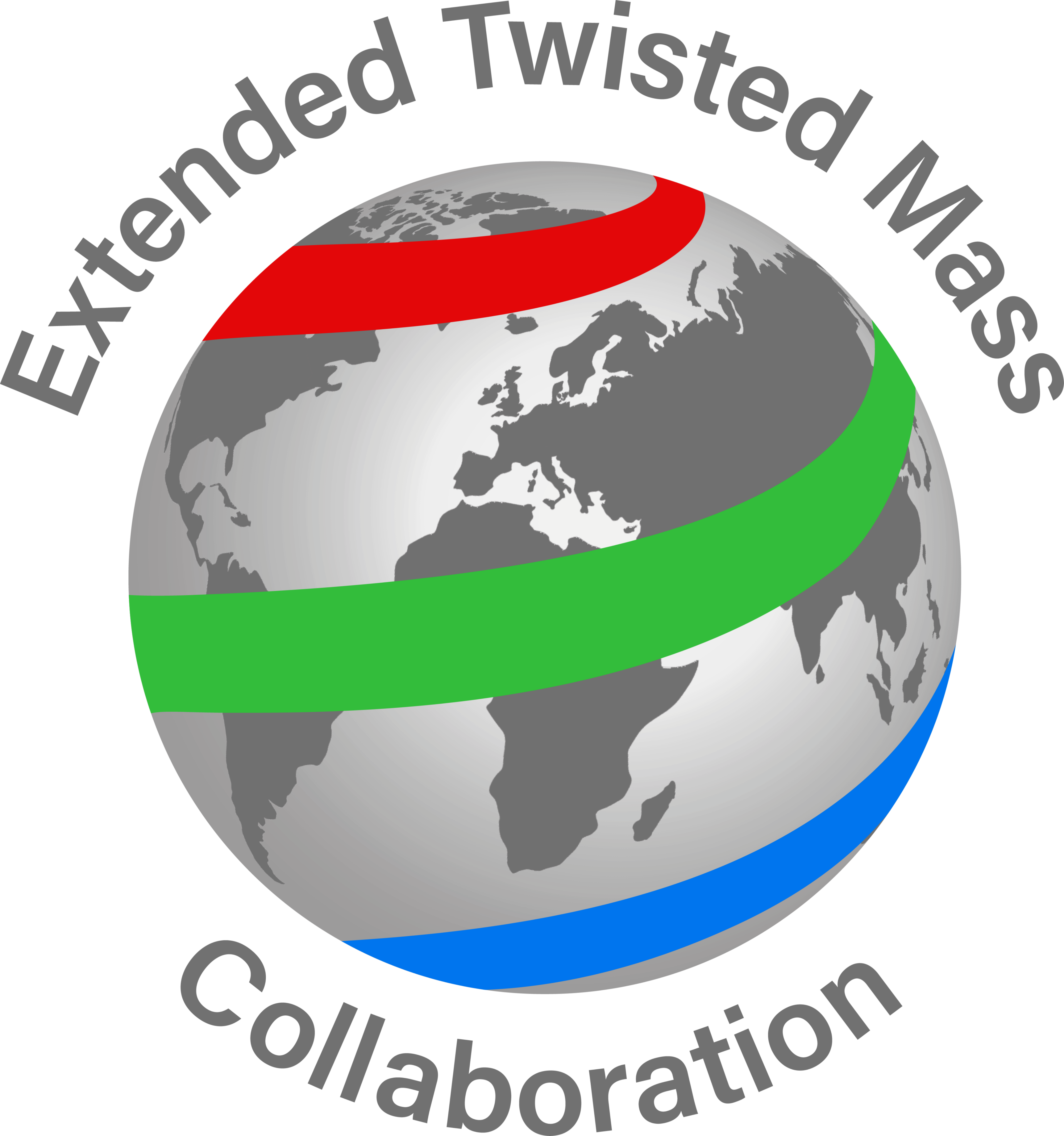}}
\vspace{0.3cm}
We present the results of a first-principles theoretical study of the inclusive semileptonic decays of the \texorpdfstring{$D_s$}{Ds} meson. We performed a state-of-the-art lattice QCD calculation by 
taking into account all sources of systematic errors. A detailed discussion  
of our lattice calculation, demonstrating that inclusive semileptonic decays can nowadays be studied on the lattice at a phenomenologically relevant level of accuracy, is the subject of a companion paper~\cite{DeSantis:2025qbb}. Here we focus on the phenomenological implications of our 
results. Using the current best estimates of the relevant Cabibbo--Kobayashi--Maskawa (CKM) matrix elements, our theoretical predictions for the decay rate and for the first two lepton-energy moments are in very good agreement with the corresponding experimental measurements. We also argue that, 
while the inclusive \texorpdfstring{$D_s$}{Ds} channel is not yet competitive with the exclusive channels in the $|V_{cs}|$ determination, the situation can be significantly improved in the near future. 
\end{abstract}

\maketitle

\section{Introduction}
The study of the flavor sector of the Standard Model (SM) of elementary particles is a topic of pivotal importance in fundamental physics. In particular, the elements of the CKM quark mixing matrix~\cite{Cabibbo:1963yz, Kobayashi:1973fv} are potentially sensitive to physics beyond the SM, because many of the proposed SM extensions lead to new flavor-changing interactions, lepton-flavor universality violations and additional sources of CP violation. 
These effects may be observable at current collider experiments, and could thus provide an exciting opportunity for (indirect) detection of new physics. In fact, in recent years various experimental collaborations have reported tensions between SM predictions and observations~\cite{BaBar:2012obs, BaBar:2013mob, LHCb:2013ghj, LHCb:2014vgu, LHCb:2015wdu, Belle:2015qfa, Bifani:2018zmi, LHCb:2021trn,Muong-2:2006rrc,Muong-2:2021ojo,Muong-2:2023cdq}. In parallel, tantalizing discrepancies between the values measured for the $|V_{ub}|$ and $|V_{cb}|$ CKM matrix elements from inclusive and exclusive decays persist~\cite{HeavyFlavorAveragingGroupHFLAV:2024ctg}. It is, therefore, very timely to improve the knowledge of flavor-changing processes associated with the elements of the CKM matrix, including, in particular, semileptonic decays of pseudoscalar mesons, which couple the leptonic and the hadronic flavor sectors.

Given the non-perturbative nature of these processes, lattice QCD represents the only ab-initio framework enabling controlled theoretical predictions of the associated decay rates. In the past few decades, lattice QCD has been used to compute a broad variety of physical quantities (see e.g. Refs.~\cite{BMW:2014pzb,RBC:2020kdj} and also Ref.~\cite{Aliberti:2025beg} and references therein) and is now a well-established toolbox to study flavor-physics from first principles~\cite{FlavourLatticeAveragingGroupFLAG:2024oxs}. Lattice QCD studies of exclusive semileptonic decays of kaons and heavy pseudoscalar mesons are relatively straightforward, and are now at or close to sub-percent precision levels. In contrast, the study of inclusive decays on the lattice has proven to be much more challenging: this is mainly due to the difficulty of taking into account a very large number of physical states, including many-hadron ones. Recently, however, various works suggested that the differential rates of inclusive semileptonic decays could be extracted from suitable correlators evaluated on the lattice~\cite{Hashimoto:2017wqo, Hansen:2017mnd, Hansen:2019idp, Bulava:2019kbi, Bulava:2021fre, Gambino:2020crt}. In particular, in Ref.~\cite{Gambino:2020crt} it was shown that these rates can be obtained by computing suitably smeared spectral densities, associated with a class of four-point Euclidean correlation functions. Even though the reconstruction of a spectral density from a finite number of values of the correlator, with their own numerical uncertainties, is non-trivial, in the past few years various different methods have been proposed to tackle this problem~\cite{Asakawa:2000tr, Meyer:2011gj, Burnier:2013nla, Rothkopf:2019ipj, Hansen:2019idp, Kades:2019wtd, Bailas:2020qmv, Horak:2021syv, Bergamaschi:2023xzx}. 

Following the feasibility study that was presented in Ref.~\cite{Gambino:2022dvu}, we used the variant of the Backus--Gilbert method~\cite{Backus:1968svk} that was put forward in Ref.~\cite{Hansen:2019idp} and performed a systematic, complete theoretical investigation of inclusive semileptonic decays of the $D_s$ meson using lattice QCD in the isospin symmetric limit\footnote{
After Ref.~\cite{Gambino:2022dvu} also another lattice group started to face the same challenge~\cite{Barone:2022gkn,Kellermann:2022mms,Barone:2023tbl,Kellermann:2023yec,Barone:2023iat,Kellermann:2024zfy,Hashimoto:2024pnd}. See also Ref.~\cite{Kellermann:2025pzt}, which appeared after the completion of this work, for a study at fixed lattice spacing and unphysical pion mass of the same process.}. A detailed discussion of all the steps of the lattice calculation, including a careful analysis of all sources of systematic errors, is presented in the companion paper~\cite{DeSantis:2025qbb}. In this letter, after introducing the problem and briefly discussing the methodological aspects of the calculation, we focus on the phenomenological implications of our results by comparing them  with the available experimental data~\cite{CLEO:2009uah,BESIII:2021duu}.

\section{Methods}
\begin{figure}[]
\includegraphics[width=0.9\columnwidth]{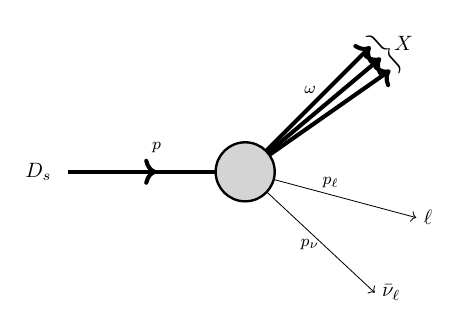}
\caption{Kinematics of the inclusive semileptonic decays of the $D_s$ meson analyzed in this work.\label{fig:kinematics}} 
\end{figure}
A schematic view of the inclusive $D_s\mapsto X \ell \bar \nu_\ell$ process is shown in \cref{fig:kinematics}, where $p$ denotes the four-momenta of the $D_s$ meson, $p_\ell$ of the outgoing lepton $\ell$, $p_\nu$ of the anti-neutrino $\bar \nu_\ell$ and $\omega$ of the generic final hadronic state $X$. For later convenience, we write $\omega=m_{D_s}(\omega_0, \vec \omega)$, with $m_{D_s}$ the mass of the $D_s$ meson.

The inclusive differential decay rate involves the squared modulus of the matrix element representing the transition from the initial $D_s$ state to the $X \ell \bar \nu_\ell$ final state. To disentangle the high-energy physics (at the electro-weak scale) from the low-energy physics probed in lattice QCD, the matrix element is written in terms of The Fermi effective Hamiltonian, involving the hadronic electroweak current
$J^\mu(x) = 
V_{cs} J^\mu_{\bar cs}(x) +
V_{cd} J^\mu_{\bar cd}(x) +
V_{us} J^\mu_{\bar us}(x)$,
which is the sum of the different flavour contributions $J_{\bar f g}^\mu(x)=\bar{\psi}_f(x) \gamma^\mu (1-\gamma^5) \psi_g(x)$ weighted by the corresponding CKM matrix elements.
One has to integrate over the three-momenta of the lepton and neutrino, and sum over all of the allowed final hadronic states, with four-momentum conservation enforced. Since, by neglecting non-factorizable electro-weak corrections, the hadronic and leptonic parts do not interfere with each other, the contribution to the differential rate (for each combination of flavors) can be written in terms of the contraction between the leptonic tensor $L^{\alpha\beta}(p_\ell,p_\nu) = 4\left( p_\ell^\alpha p_\nu^\beta + p_\ell^\beta p_\nu^\alpha - g^{\alpha\beta} p_\ell\cdot p_\nu +i\epsilon^{\alpha\beta\gamma\delta} (p_\ell)_\gamma (p_\nu)_\delta \right)$ and the hadronic tensor
\begin{align}
\label{hadronic_tensor}
H_{\mu\nu}(p,\omega)=\frac{(2\pi)^4}{2m_{D_s}} \bra{D_s(p)}J_\mu^\dagger(0)\, \delta^4(\mathbb{P}-\omega)\, J_\nu(0)\ket{D_s(p)},
\end{align}
where $\mathbb{P}$ is the QCD four-momentum operator. Using special relativity, $H$ can be decomposed in terms of five independent spectral-densities form-factors which, in the reference-frame of the $D_s$ meson, are functions of $\omega_0$ and of $\vec \omega^2$. From these, one can construct three linear combinations, $Z^{(0)}$, $Z^{(1)}$, and $Z^{(2)}$, that enter the expression of the differential decay rate integrated over the lepton energy. Denoting the upper integration-limit for $\omega_0$ as $\omega^\mathrm{max}$, one has
\begin{align}
\label{in_terms_of_Zi}
&
\frac{1}{\bar{\Gamma}}\,
\frac{d \Gamma}{d \vec \omega^2} = \frac{1}{\bar{\Gamma}}\sum_{p=0}^2 \frac{d\Gamma^{(p)}}{d \vec \omega^2} =
\nonumber \\
&= 
\sum_{p=0}^2 | \vec \omega |^{3-p} 
\int_0^{\omega^\mathrm{max}}d\omega_0\,
(\omega^\mathrm{max}-\omega_0)^p\, Z^{(p)}(\omega_0, \vec \omega^2),
\end{align}
where $\bar{\Gamma} = m_{D_s}^5 \GF^2 \Sew/(48\pi^4)$, $\GF=1.1663788(6)\times 10^{-5}$~GeV$^{-2}$~\cite{ParticleDataGroup:2024cfk} is the Fermi constant and  $\Sew=1.013$  accounts for the logarithmic electroweak correction~\cite{Sirlin:1981ie} and for QED threshold corrections\footnote{QED threshold corrections are different in the $\bar u s$ channel. However, in~\cite{DeSantis:2025qbb} we show that $\Gamma_{\bar u s}$ is negligible at the current level of accuracy. } \cite{Bigi:2023cbv}. 
 
It is convenient to express the integral of \cref{in_terms_of_Zi} as
\begin{align}
\label{in_terms_of_Zi_integral}
\lim_{\sigma\mapsto 0}
\int_0^\infty d\omega_0
\Theta^{(p)}_\sigma(\omega^\mathrm{max}-\omega_0) Z^{(p)}(\omega_0, \vec \omega^2),
\end{align}
where the kernels $\Theta^{(p)}_\sigma(x)=x^p \Theta_\sigma(x)$ are obtained by introducing a ``smoothed'' version of the Heaviside function $\theta(x)$, depending on a smearing parameter $\sigma$ and such that $\lim_{\sigma\mapsto 0}\Theta_\sigma(x) = \theta(x)$. Indeed, the hadronic tensor of \cref{hadronic_tensor}, and therefore its linear combinations $Z^{(p)}(\omega_0, \vec \omega^2)$, cannot be directly computed on the lattice. It is possible, instead, to compute the (amputated) correlator $ \bra{D_s(p)} T\{J_\mu^\dagger(t,\vec x)\, J_\nu(0)\}\ket{D_s(p)}$ at discrete values of the Euclidean times $t=an$, where $a$ is the lattice spacing. From this~\cite{Gambino:2020crt,DeSantis:2025qbb}, by taking the required linear combinations, one gets the three correlators 
\begin{flalign}
\hat Z^{(p)}(t,\vec \omega^2) = \int_0^\infty d\omega_0\, e^{-m_{D_s}\omega_0t}\, Z^{(p)}(\omega_0,\vec \omega^2)\;,
\label{Zcorr}
\end{flalign}
which are in fact the Laplace transform of the $Z^{(p)}(\omega_0, \vec \omega^2)$ distributions.
The calculation of the $D_s$ inclusive decay rate $\Gamma$  is then reduced to the problem of establishing a connection between \cref{in_terms_of_Zi_integral} and \cref{Zcorr}. 

The problem can be addressed in a mathematically well-defined, systematically improvable way through the Hansen-Lupo-Tantalo (HLT) method of Ref.~\cite{Hansen:2019idp}. The basic idea consists in relying on the Stone-Weierstrass theorem  to represent the smooth kernels of \cref{in_terms_of_Zi_integral} according to
\begin{flalign}
\label{eq:HLTkernel}
\Theta^{(p)}_\sigma(\omega^\mathrm{max}-\omega_0)
=
\lim_{N\mapsto \infty}\sum_{n=1}^N g^{(p)}_n(N)\, 
e^{-am_{D_s}\omega_0\, n}\;,
\end{flalign}
i.e. on the basis-functions given by the exponentials appearing in \cref{Zcorr} with coordinates given by the vector of the coefficients $\vec g^{(p)}$. These coefficients are fixed by minimizing a linear combination of the so-called ``norm'' and ``error'' functionals. The norm functional measures the distance between the exact kernel $\Theta^{(p)}_\sigma(\omega^\mathrm{max}-\omega_0)$ and its approximation according to the r.h.s.\ of \cref{eq:HLTkernel} at finite $N$. The error functional, originally introduced by  Backus and Gilbert~\cite{Backus:1968svk}, is defined in terms of the covariance matrix of the correlators $\hat Z^{(p)}$  evaluated at finite lattice spacing. Thus, the HLT algorithm improves on the classical Backus-Gilbert method by using  a desired smearing kernel as input, and is designed to provide an optimal trade-off between the ``accuracy'' and the ``precision'' of the numerical solution to the problem.  

By using the coefficients $\vec g^{(p)}$ obtained through the HLT algorithm, the differential decay rate can be obtained according to
\begin{align}
\label{in_terms_of_hatZi}
\frac{1}{\bar{\Gamma}}\,
\frac{d \Gamma}{d \vec \omega^2} = \sum_{p=0}^2 | \vec \omega |^{3-p} 
\lim_{\sigma\mapsto 0}
\lim_{N\mapsto \infty}
\sum_{n=1}^N g^{(p)}_n(N)\, 
\hat Z^{(p)}(an,\vec \omega^2)\;.
\end{align}
Similar formulae can be introduced for the leptonic moments $M_n$ (defined by weighting the differential decay rate with the $n-$th power of the lepton energy and normalizing by the total rate) and the associated differential quantities $dM_n/d \vec \omega^2$ (see Sec. IV of Ref.~\cite{DeSantis:2025qbb}).

\begin{figure}[!th]
\centering
\includegraphics[width=\columnwidth]{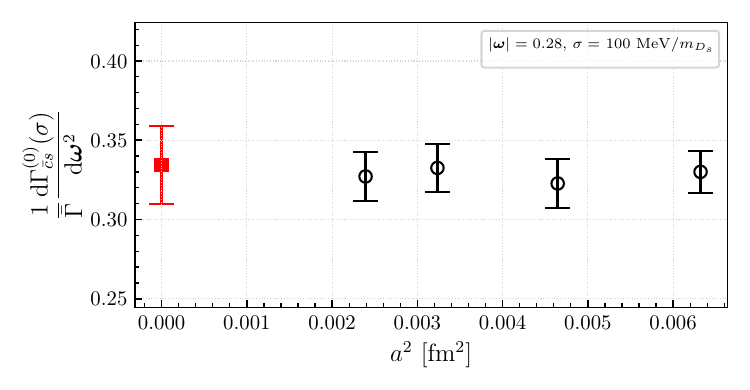}
\includegraphics[width=\columnwidth]{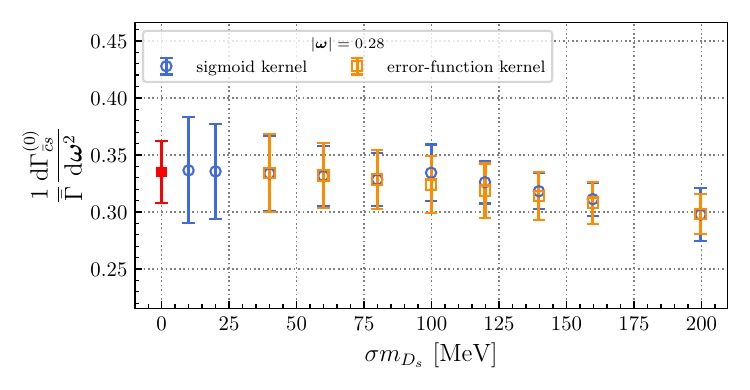}
\caption{ Examples of continuum ($a\mapsto 0$), and $\sigma \mapsto 0$ extrapolations of $\bar{\Gamma}^{-1} d\Gamma^{(0)}/d\vec{\omega}^2$ in the $\bar{c}s$ channel at $|\vec{\omega}|=0.28$. Top: data at four lattice spacings (black points) for $\sigma m_{D_{s}} = 100\,\mathrm{MeV}$ and the continuum-extrapolated value (red). 
Bottom: the red point represents the result of the $\sigma \mapsto 0$ extrapolation based on a joint fit of the continuum limit results obtained using the so-called ``sigmoid'' (blue points) and ``error-function'' (orange points) representations of $\Theta_\sigma(x)$, defined at the beginning of Section VII of the companion paper \cite{DeSantis:2025qbb}.
For details on the procedures used to perform the two limits, see Sec.~VIII-A of Ref.~\cite{DeSantis:2025qbb}.
\label{fig:limits}}
\end{figure}
We have computed the relevant Euclidean correlators on a set of ensembles of lattice QCD gauge field configurations produced by the Extended Twisted Mass Collaboration (ETMC)~\cite{Alexandrou:2018egz, ExtendedTwistedMass:2020tvp, ExtendedTwistedMass:2021qui, Finkenrath:2022eon}: they were generated including the  dynamical effects of two (mass-degenerate) light quark flavors, the strange quark, and the charm quark, using the Wilson-Clover Twisted-Mass discretization for fermionic fields~\cite{Frezzotti:2000nk,Frezzotti:2003xj,Frezzotti:2004wz}, at four values of the lattice spacing ($a\ge 0.049$~fm) and three values of the physical volume ($L\le 7.6$~fm).

In the companion paper~\cite{DeSantis:2025qbb} we provide 
a detailed description of all the steps of the numerical calculation, including a thorough discussion of the specific implementation of the HLT algorithm. Here, before discussing the phenomenological implications of our results, we illustrate in \cref{fig:limits} the quality of the data we use for the continuum and $\sigma\mapsto 0$ extrapolations of $\bar{\Gamma}^{-1}d\Gamma^{(0)}/d\vec{\omega}^2$ in the dominant $\bar{c}\,s$ channel at $|\vec{\omega}|=0.28$. 
The top panel shows results for $\sigma = 100\,\mathrm{MeV}/m_{D_s}$ at four values of the lattice spacings (black points) and the continuum-extrapolated value (red point), obtained by combining multiple fits using the Bayesian Model Average approach in Ref.~\cite{DeSantis:2025qbb}. The bottom panel demonstrates the $\sigma \mapsto 0$ limit for two different definitions of the kernel $\Theta_\sigma(x)$ (blue and orange points), which coincide as $\sigma \mapsto 0$. The red point is the result of a combined $\sigma \mapsto 0$ fit guided by the theoretical asymptotic formulae derived in Ref.~\cite{DeSantis:2025qbb}.

\section{Results}
\begin{table}[t]
\begin{tabular}{lcc}
$\bar f g$ & $\quad\qquad\bar c s\quad\qquad$ & $\quad\qquad\bar c d\quad\qquad$ \\ [4pt] 
\hline
\\
 $10^{14}\times\Gamma_{\bar f g}$ [GeV]               & $ 8.53(46)(30)[55]$  & $ 12.60(79)(49)[93]$ \\ [8pt]
 $\Gamma M_{1,\bar f g}/\Gamma_{\bar cs}$ [GeV]      & $ 0.453(21)(11)[24]$ & $ 0.731(53)(30)[61]$   \\ [8pt]
 $\Gamma M_{2,\bar f g}/\Gamma_{\bar cs}$ [GeV$^2$]  & $ 0.223(9)(6)[11]$ & $ 0.416(37)(22)[43] $  \\ [2pt]
\end{tabular}
\caption{Our final determinations of the decay rate and the first two lepton-energy moments for the two dominating channels. The covariance matrix associated with these results is given in \cref{tab:covariance_results}.
\label{tab:results}}
\end{table}
\begin{figure}[!tb]
\centering
\includegraphics[width=\columnwidth]{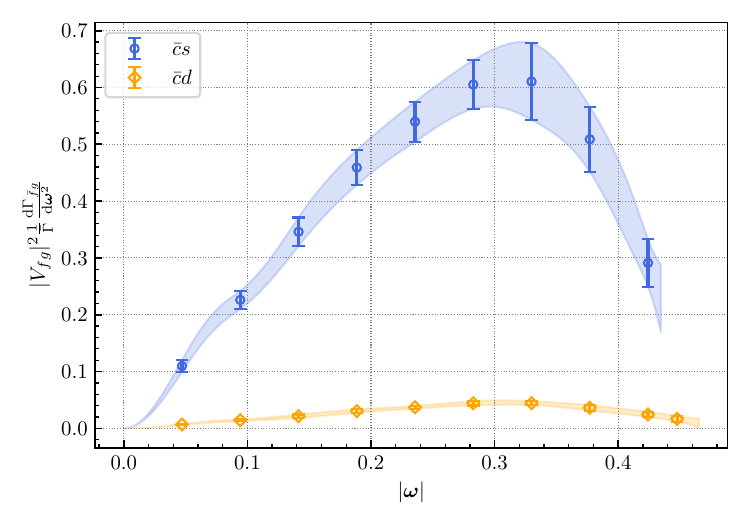}
\includegraphics[width=\columnwidth]{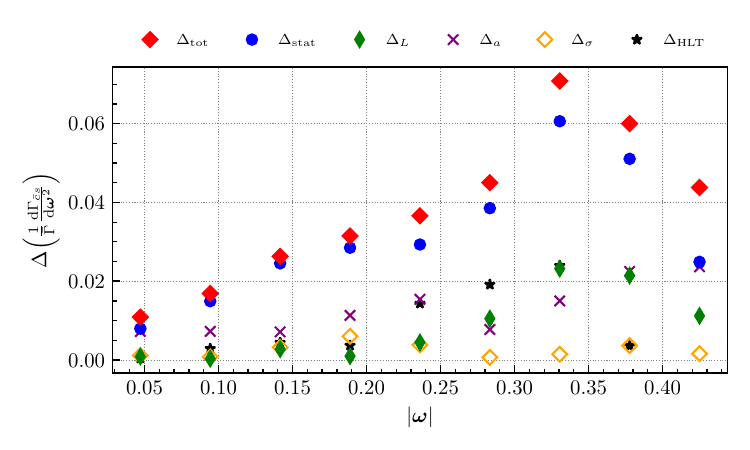}
\includegraphics[width=\columnwidth]{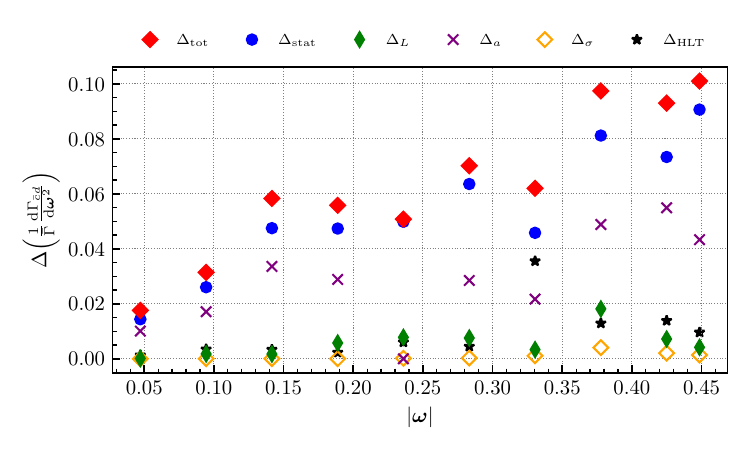}
\caption{{\textit{Top}}: contributions to the differential decay rate of the dominant $\bar{c}s$ and $\bar{c}d$ channels (multiplied by the squared moduli of the corresponding CKM elements), shown together with a cubic-spline interpolation to the simulated momenta. {\textit{Middle/Bottom}}: error budgets for the $\bar{c}s$ and $\bar{c}d$ channels, respectively. Red and blue points represent the total ($\Delta_\mathrm{tot}$) and statistical ($\Delta_\mathrm{stat}$) errors, while the green, purple, yellow, and black points represent the error associated with the infinite-volume ($\Delta_L$), continuum-limit ($\Delta_a$), $\sigma\mapsto 0$ ($\Delta_\sigma$) extrapolation, and with the HLT-reconstruction ($\Delta_\mathrm{HLT}$), respectively. The dominant source of error is statistical.}
    \label{fig:final_differential_DGammaDq2}
\end{figure}

The total decay rate $\Gamma$ is obtained by adding the integrals over ${\vec \omega}^2$ of the differential decay rates obtained in the different flavour channels. 
In the companion paper~\cite{DeSantis:2025qbb} we have computed 
the $\Gamma_{\bar u s}$ contribution to the total rate and shown that, at the present level of accuracy, it is negligible w.r.t.\ the statistical uncertainties of the dominating  $\Gamma_{\bar c s}$ and 
$\Gamma_{\bar c d}$ contributions. We have also discussed how to interpolate and integrate the results for the ${\bar{c} s}$ and ${\bar{c} d}$ flavor channels that are shown in \cref{fig:final_differential_DGammaDq2} together with the associated error-budgets. The calculation of the first two leptonic moments is carried out in a similar way.

Our final results are reported in \cref{tab:results} and, by using the PDG~\cite{ParticleDataGroup:2024cfk} values $|V_{cs}|^\mathrm{PDG}=0.975(6)$ and $|V_{cd}|^\mathrm{PDG}=0.221(4)$, we get
\begin{align}
\label{Gamma_results}
&
\Gamma
= 
|V_{cs}|^2\, \Gamma_{\bar c s}
+
|V_{cd}|^2\, \Gamma_{\bar c d}
=
8.72(47)(31)[56]\times 10^{-14}\ \mathrm{GeV}\;,
\nonumber \\[8pt]
&
M_1= 
\frac{ 
\frac{\Gamma M_{1,\bar c s}}{\Gamma_{\bar c s}}
+
\frac{|V_{cd}|^2}{|V_{cs}|^2\,}\, \frac{\Gamma M_{1,\bar c d}}{\Gamma_{\bar c s}}
}
{
1
+
\frac{|V_{cd}|^2}{|V_{cs}|^2\,}\, \frac{\Gamma_{\bar c d}}{\Gamma_{\bar c s}}
}
=
0.456(19)(11)[22]\, \mathrm{GeV}\;,
\nonumber \\[8pt]
&
M_2= 
\frac{ 
\frac{\Gamma M_{2,\bar c s}}{\Gamma_{\bar c s}}
+
\frac{|V_{cd}|^2}{|V_{cs}|^2\,}\, \frac{\Gamma M_{2,\bar c d}}{\Gamma_{\bar c s}}
}
{
1
+
\frac{|V_{cd}|^2}{|V_{cs}|^2\,}\, \frac{\Gamma_{\bar c d}}{\Gamma_{\bar c s}}
}
=
0.227(9)(5)[10]\, \mathrm{GeV}^2\;.
\end{align}

\begin{figure}[]
\centering
\includegraphics[width=\columnwidth]{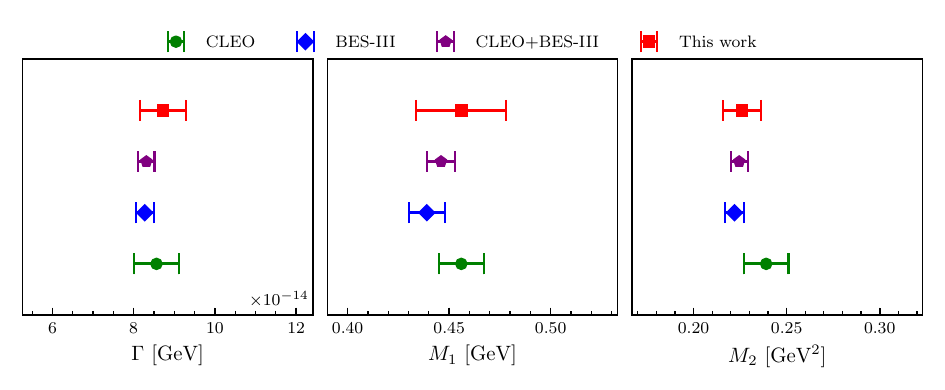}
\caption{Comparison between the experimental results from Refs.~\cite{CLEO:2009uah, BESIII:2021duu, ParticleDataGroup:2024cfk,Gambino:2010jz} and our theoretical prediction (red points), for the decay rate (left-panel) and for the first (middle-panel) and second (right-panel)  lepton moment.\label{fig:final_results}}
\end{figure}

Our results can be compared with experimental measurements from the CLEO~\cite{CLEO:2009uah} and BES-III~\cite{BESIII:2021duu} collaborations, as well as their combined average~\cite{ParticleDataGroup:2024cfk}: 
\begin{align}
& \Gamma^\mathrm{CLEO} = 8.56(55)\times 10^{-14}\,\mathrm{GeV},\\
& \Gamma^\mathrm{BES-III}=8.27(22 )\times 10^{-14}\,\mathrm{GeV},\\
& \Gamma^\mathrm{average} = 8.31(20)\times 10^{-14}\,\mathrm{GeV},
\end{align}
which are in excellent agreement with our theoretical prediction in \cref{Gamma_results}, as shown in the left panel of \cref{fig:final_results}.
For the first two leptonic moments, the experimental results by the CLEO and BES-III collaborations are:
\begin{align}
&
M_1^\mathrm{CLEO}  = 0.456(11) \,\mathrm{GeV},\\
&
M_1^\mathrm{BES-III}  = 0.439(9) \,\mathrm{GeV},\\
&
M_1^\mathrm{average}  = 0.446(7) \,\mathrm{GeV},
\end{align}
for $M_1$, and
\begin{align}
&
M_2^\mathrm{CLEO} = 0.239(12) \,\mathrm{GeV}^2,\\
&
M_2^\mathrm{BES-III} = 0.222(5) \,\mathrm{GeV}^2,\\
&
M_2^\mathrm{average} = 0.2245(46) \,\mathrm{GeV}^2,
\end{align}
for $M_2$. We have obtained these results by repeating also in the case of the BES-III data the analysis performed in Ref.~\cite{Gambino:2010jz} in the case of the CLEO results. Like for the decay rate, the agreement of our theoretical predictions of \cref{Gamma_results} with the experimental results is excellent, as the middle- and right-panel of \cref{fig:final_results} show.

A complementary analysis that can be carried out is to convert the comparison between lattice and experiments  into a determination of the CKM matrix elements. In principle, the relations in \cref{{Gamma_results}} about $M_1$ and $M_2$, after setting the left-hand side respectively to $M_1^{\mathrm{average}}$ or $M_2^{\mathrm{average}}$, can be solved for the ratio $|V_{cd}|^2 / |V_{cs}|^2$. However, since for both $l=1,2$ the $\Gamma M_{l,\bar{c}s}/\Gamma_{\bar c s}$ contribution alone (which is CKM-independent) already agrees with the experimental result $M_{l}^\mathrm{average}$ within uncertainties (see~\cref{tab:results}), there is essentially no remaining ``room'' for extracting $|V_{cd}|^2 / |V_{cs}|^2$ at better than the $100\%$-level of precision. In other words, the measured moments are entirely saturated, within errors, by the CKM-independent part of the theoretical predictions. This also implies that, currently, the comparison between lattice and experiments for these two observables remains essentially unaffected by moderate variations in the CKM parameters. Therefore, since this is the very first time that lattice QCD confronts experiments on these quantities, the observed agreement is remarkable.

As for the total decay rate in Eq.~(\ref{Gamma_results}), we have solved Eq.~(\ref{Gamma_results}) by setting its left-hand side to $\Gamma^{\rm average}$ while fixing $|V_{cd}|$ to  $|V_{cd}|^{\rm PDG}$. This allows us to extract $|V_{cs}|$, for which we obtain
\begin{align}
|V_{cs}| = 0.951\,(35)~,
\end{align}
in very good agreement with $|V_{cs}|^{\rm PDG} = 0.975(6)$. The determination of $|V_{cs}|$ obtained this way is largely independent on the specific value of $|V_{cd}|$ used. Varying $|V_{cd}|^2$ even 
by $50\%$ relative to its PDG value changes $|V_{cs}|$ by less than $0.5\,\sigma$. The current accuracy of our $|V_{cs}|$ determination is at the level of $\simeq 3.5\%$, which has not yet reached the precision found in exclusive semileptonic decays. The larger uncertainties stem from both the lattice computation (mainly statistical errors, see \cref{fig:final_differential_DGammaDq2}) and, to a slightly smaller extent, experimental measurements. Both of them can~\cite{Bernlochner:2024vhg} (and will) be significantly improved in the coming years: the $1\%$ precision goal in isospin symmetric QCD is well within reach.

We conclude this section with another interesting observation. In the $\bar{c}d$ channel, we can compare our (fully inclusive) result $\Gamma_{\bar{c}d} = 0.62(5) \times 10^{-14}~\mathrm{GeV}$
with the sum of the available exclusive-channel measurements, to check for potentially missing contributions. Measurements of the branching fractions exist for $D_{s}^{+} \mapsto K^{0} e^{+} \nu_{e}$ and $D_{s}^{+} \mapsto K^{*0} e^{+} \nu_{e}$~\cite{Hietala:2015jqa,BESIII:2018xre}, yielding
$\Gamma_{\bar{c}d}^{K^{0}+K^{*0}} = 0.722(66)\times 10^{-14}~{\rm GeV}$. This result is compatible, at the $1.2\,\sigma$ level, with our inclusive result, indicating that at the current level of accuracy, the $K^{0}$ and $K^{*0}$ contributions saturate the inclusive rate $\Gamma_{\bar{c}d}$.

\section{Conclusions}
In this letter, and the companion paper~\cite{DeSantis:2025qbb}, we presented Standard Model predictions for the inclusive semileptonic decay rate of the $D_s$ meson and its first two lepton-energy moments. Our \textit{ab-initio} approach starts from the SM Lagrangian, includes the dominant higher-order electroweak corrections,
and treats QCD non-perturbatively on the lattice, without any uncontrolled approximation. By combining recent formal advances to deal with the large number of hadronic final states involved~\cite{Gambino:2020crt} and state-of-the-art spectral-reconstruction techniques~\cite{Hansen:2019idp}, we successfully computed these observables by taking into accout all sources of systematic errors. Furthermore, for the first time in the case of inclusive semileptonic decays, we performed a comparison between first-principles lattice QCD predictions and experimental results.

By using the PDG~\cite{ParticleDataGroup:2024cfk} values of the CKM matrix elements, our final results for $\Gamma$, $M_1$, and $M_2$ in \cref{Gamma_results} are in very good agreement with experiments. At the same time, we observe that at the current level of theoretical and experimental accuracy, this channel is not sensitive to moderate variations of $|V_{cd}|$. Therefore, we used our theoretical predictions and $|V_{cd}|^\mathrm{PDG}$ to obtain a determination of $|V_{cs}|$ with $O(3\%)$ total accuracy, in perfect agreement with the current PDG determination. In fact, at present, the inclusive $D_s\mapsto X \ell \bar \nu_\ell$ channel is not competitive with  more precise exclusive determinations of $|V_{cs}|$. On the other hand, since 
our theoretical error is dominated by the statistical uncertainty and 
the experimental errors can likely be reduced  \cite{Bernlochner:2024vhg}, the situation will certainly improve in the future.

As an extension of the present study, one obvious choice would be to carry out a similar analysis for $B$ mesons. Indeed, the results presented in this letter and in the companion paper~\cite{DeSantis:2025qbb} demonstrate that inclusive semileptonic decays of heavy mesons can now be studied on the lattice from first-principles at a phenomenologically relevant level of accuracy. We are going to exploit the implications of this exciting new perspective in a future work~\cite{DeSantis_et_al_Bmeson}.

\section{Acknowledgements}

The authors gratefully acknowledge the Gauss Centre for Supercomputing e.V. (www.gauss-centre.eu) for funding this project by providing computing time on the GCS Supercomputer JUWELS~\cite{JUWELS} at Jülich Supercomputing Centre (JSC) and on the GCS Supercomputers SuperMUC-NG at Leibniz Supercomputing Centre, and the granted access to the Marvin cluster hosted by the University of Bonn. The authors acknowledge the Texas Advanced Computing Center (TACC) at The University of Texas at Austin for providing HPC resources (Project ID PHY21001). The authors gratefully acknowledge PRACE for awarding access to HAWK at HLRS within the project with Id Acid 4886. We acknowledge the Swiss National Supercomputing Centre (CSCS) and the EuroHPC Joint Undertaking for awarding this project access to the LUMI supercomputer, owned by the EuroHPC Joint Undertaking, hosted by CSC (Finland) and the LUMI consortium through the Chronos programme under project IDs CH17-CSCS-CYP. We acknowledge EuroHPC Joint Undertaking for awarding the project ID EHPC-EXT-2023E02-052 access to MareNostrum5 hosted by at the Barcelona Supercomputing Center, Spain.
This work has been supported  by the MKW NRW under the funding code NW21-024-A as part of NRW-FAIR and by the Italian Ministry of University and Research (MUR) and the European Union (EU) – Next Generation EU, Mission 4, Component 1, PRIN 2022, CUP F53D23001480006 and CUP D53D23002830006.
We acknowledge support from the ENP, LQCD123, SFT, and SPIF Scientific Initiatives of the Italian Nuclear Physics Institute (INFN). F.S. is supported by ICSC-Centro Nazionale di Ricerca in High Performance Computing, Big Data and Quantum Computing, funded by European Union-NextGenerationEU  and by Italian Ministry of University and Research (MUR) projects FIS 0000155. A.S. is supported by STFC grant ST/X000648/1. We thank Paolo Garbarino for helpful discussions.

\begin{widetext}

\begin{table}[t]
\begin{tabular}{lcccccc}
  & \quad $10^{14}\times\Gamma_{\bar c s}$ \quad 
  & \quad $10^{14}\times\Gamma_{\bar c d}$ \quad 
  & \quad $\Gamma M_{1,\bar c s}/\Gamma_{\bar cs}$ \quad 
  & \quad $\Gamma M_{1,\bar c d}/\Gamma_{\bar cs}$ \quad 
  & \quad $\Gamma M_{2,\bar c s}/\Gamma_{\bar cs}$ \quad 
  & \quad $\Gamma M_{2,\bar c d}/\Gamma_{\bar cs}$\\ [4pt] 
\hline
\\
\quad $10^{14}\times\Gamma_{\bar c s}$         & 30.783  &  23.542  &   0.139  &  -0.890  &  -0.004  &  -0.505      \\ [8pt]
\quad $10^{14}\times\Gamma_{\bar c d}$         & 23.542  &  85.857  &  -0.037  &   3.044  &  -0.067  &   1.689      \\ [8pt]
\quad $\Gamma M_{1,\bar c s}/\Gamma_{\bar cs}$ & 0.139  &  -0.037  &   0.055  &  -0.002  &   0.019  &   0.003      \\ [8pt]
\quad $\Gamma M_{1,\bar c d}/\Gamma_{\bar cs}$ & -0.890  &   3.044  &  -0.002  &   0.378  &   0.003  &   0.233      \\ [8pt]
\quad $\Gamma M_{2,\bar c s}/\Gamma_{\bar cs}$ & -0.004  &  -0.067  &   0.019  &   0.003  &   0.011  &   0.006      \\ [8pt]
\quad $\Gamma M_{2,\bar c d}/\Gamma_{\bar cs}$ & -0.505  &   1.689  &   0.003  &   0.233  &   0.006  &   0.181      \\ [2pt]
\end{tabular}
\caption{Covariance matrix of the results given in \cref{tab:results}. All the numbers in the table are multiplied by $10^2$.
\label{tab:covariance_results}}
\end{table}

\end{widetext}

\FloatBarrier

\bibliography{short_incl_01}
\end{document}